\UseRawInputEncoding
\documentclass[12pt,letterpaper]{article}
\usepackage{amsmath,amssymb,pgf,pgfarrows,pgfnodes,float,appendix, hyperref}
\usepackage{graphicx}
\usepackage{subfigure}
\usepackage[margin=0.9in]{geometry}

\newcommand{\be}{\begin{equation}}
\newcommand{\ee}{\end{equation}}
\newcommand{\bea}{\begin{eqnarray}}
\newcommand{\eea}{\end{eqnarray}}

\title{{\rm\footnotesize \qquad \qquad \qquad \qquad \qquad \ \qquad \qquad \qquad \ \ \ \ \ \                 RUNHETC-2025-36}\vskip.5in   Finite Entropy Implies Finite Dimension in Quantum Gravity}
\author{Tom Banks\\
Department of Physics and NHETC\\
Rutgers University, Piscataway, NJ 08854\\
E-mail: \href{mailto:tibanks@ucsc.edu}{tibanks@ucsc.edu}
\\
\\
\\
\\
}
\date{}
\begin{document}
\maketitle

\begin{abstract}  Quantum Field Theory (QFT) introduced us to the notion that a causal diamond in space-time corresponded to a subsystem of a quantum mechanical system defined on the global space-time. Work by Jacobson\cite{ted95}, Fischler and Susskind\cite{fs} and particularly Bousso\cite{bousso} suggested that in the quantum theory of gravity this subsystem should have a density matrix of finite entropy.  These authors formalized older intuitive arguments based on black hole physics.  Although mathematically, Type II von Neumann algebras admit finite entropy density matrices, the black hole arguments suggest that the number of physical states in these subsystems is finite.  The conjecture that de Sitter (dS) space has a finite number of physical states was first made in\cite{tb}\cite{wf}.  ``String theory constructions" of ``meta-stable dS states" always involve a model with an infinite number of states. The real challenge of such models is to define the asymptotic correlation functions of the space-time to which the dS space decays, and show how to isolate properties of the dS ``resonance" from those correlators.  We argue that the theory of the resonance itself is adequately described, on time scales shorter than its lifetime, by a model with a finite number of states.  

 \end{abstract}
\maketitle

\section{Causal Diamonds and Entropy}

Since the advent of local QFT, we have become used to associating causal diamonds in space-time with subsystems of the QFT Hilbert space.  This is true in the sense that they possess a von Neumann sub-algebra with non-trivial commutant in the full operator algebra, but it is well known that these algebras are Type $III_1$ and have no density matrices.  The ideas of Bekenstein-Hawking-Jacobson-Fischler-Susskind and Bousso, which associate finite entropy to these regions in a hypothetical theory of quantum gravity show us that something drastically different must happen in such a theory.  A very old hand waving argument\footnote{I first heard this from G. 't Hooft in the 1980s, but I'm sure it pre-dates that.} shows that in the semi-classical approximation most of the states that even a cut-off QFT assigns to a diamond back-react on the geometry to form a black hole larger than the diamond.  Eliminating such states leaves over a QFT entropy that scales like $(A/4G_N)^{\frac{d - 1}{d}}$\footnote{Cohen, Kaplan and Nelson\cite{CKNetal} showed that eliminating these states was entirely compatible with the experimental success of QFT.}.   This is not even enough to account for the entropy that Bekenstein and Hawking assign to a black hole and Jacobson (effectively) assigns to any causal diamond.  Indeed, 't Hooft's proposal to model black hole entropy in terms of near horizon QFT states ran into trouble\cite{thooftisrael} with back reaction long before the firewall problem proved it could not work\cite{firewall}.  

A lot of recent work, following the beautiful papers of\cite{LL} has been based on the proposal\cite{WittenLL} that one could somehow systematically correct those results by using the {\it crossed product} construction of a Type II algebra from the Type III QFT algebra.  In\cite{WittenLL}, the crossed product was taken with the action of the boundary Hamiltonian, which\cite{LL} had shown did not act on the LL algebra at leading order in $1/N$.  Indeed, that lack of action was the reason the subalgebra of a causal diamond has a non-trivial commutant.  There are a few disturbing things about this proposal, whose purpose was to construct an algebra with a trace so that entropy formulas made sense.  The most striking is that the crossed product algebra contains the original Type III algebra as a subalgebra, whereas we expect the effect of quantum gravity is to reduce the number of observables in a finite region.  

The original idea of the crossed product came from the existence of the modular group of automorphisms of a Type $III_1$ von Neumann algebra $\Delta^{it}$ and the fact that the modular operator $\Delta = e^{-K}$ had many of the properties of a density matrix, but was not an operator in the algebra itself.  Physically, the reason that diamond subalgebras in QFT do not have density matrices is because of the infinite entanglement of the diamond with states exterior to the diamond but arbitrarily close to the diamond boundary.  This has nothing to do with gauge invariance.  It is an issue in theories with no gauge invariance or in the algebra of local gauge invariant operators in Yang-Mills theory.  If we look at the sector of the gauge theory not invariant under local gauge transformations, we have an additional issue of entanglement with gauge degrees of freedom on the boundary.  The crossed product construction adds enough elements to the algebra such that density matrices like $e^{-K}$ become elements of the algebra.  In a gauge theory this should include the non-local fields that implement the entanglement with the boundary.  The authors of\cite{HKLL} have proposed a method of doing this systematically, in a particular gauge, to all orders in the $1/N$ expansion.  The proposal of\cite{WittenLL} appears to mix this up with the UV entanglement of the boundary field theory.  In a correct model of quantum gravity that UV entanglement should simply disappear.  The work of\cite{thooftisrael}\cite{firewall} shows that even a simple cutoff on the near horizon states in field theory is not the right way to treat it.  Indeed, we have known for some time\cite{swingle}\cite{harlowetal} that the proper way to treat entanglement of the bulk with the boundary in AdS/CFT is through the tensor network/Error Correcting Code formalism.  This inevitably introduces a UV cutoff length of order the AdS radius on the bulk field theory.  The interpretation of this formalism in terms of causal diamonds appeared in\cite{tbwfads1} and we are beginning to understand its connection to the work of\cite{LL}\cite{tbdoublescale}\footnote{The arguments of\cite{LL} take place in the strict $N = \infty$ limit.  However, to avoid confusion with the Type III operator algebras in the boundary CFT for finite $N$, the authors argue that they are working with a finite ultraviolet cutoff.  If one chooses that cutoff to be a lattice model of the sort used in tensor network constructions then on small enough lattices the arguments of\cite{LL}, which invoke the infinite $N$ bulk generalized free field algebras of\cite{BDHM}\cite{HKLL} are not even approximately correct.  In\cite{tbdoublescale} we speculated that the correct framework for the work of\cite{LL} was a double scaling limit where the tensor network shell is taken to the boundary as $N$ is taken to infinity. }.  

In the current note, we want to present a simple argument based on semi-classical black hole physics that the number of states in a black hole or any other finite area causal diamond in Minkowski or AdS space, is finite.  To make this argument we want to start from a model of quantum gravity that everyone believes in, and which also has no serious infrared divergence problems.  Imagine then some compactification of M theory to a Minkowski space of six or more dimensions, preserving some amount of Supersymmetry.   One can obtain such models from a non-perturbatively well defined model by taking the large radius limit of a variety of $AdS_3/CFT_2$ examples, or if one is sufficiently liberal minded, as a large $N$ limit of BFSS matrix theory.  Alternatively one can accept that many more such models are well defined to all orders in string perturbation theory and symmetry arguments prevent the appearance of non-perturbative instabilities on the moduli space.  All of these models have compactification moduli that can be chosen larger than the 11 dimensional Planck scale and therefore stable particles formed by wrapped BPS branes, which are heavy, but lighter than the Planck scale in the non-compact directions.  

We then make the following assumptions
\begin{itemize}

\item The Polchinski-Susskind\cite{polchsuss} idea that we can study Minkowski space physics (at least in $d \geq 6$ dimensions) by taking limits of CFT correlators ``focussed on an {\it arena} causal diamond in AdS space" as $(R_{AdS}/L_P) \rightarrow \infty$ with various kinematic invariants held fixed" is basically correct.

\item The arena and subregions of it can be considered quantum subsystems of a larger causal diamond, with radius a few times $R_{AdS}$.

\item The maximum number of states in that larger causal diamond is bounded by the stable AdS black hole that fills it.

\item A finite temperature state of a CFT compactified on a sphere can be calculated in a lattice theory with a finite dimensional Hilbert space at each lattice site.

\end{itemize}

Consider at some very early time a large finite number of CFT operators acting on the vacuum.  Each of these operators creates a state with the quantum numbers of a BPS brane wrapped around some cycle in the compact directions whose radius is independent of the AdS radius.  The simplest examples of models where this is possible are those of the form $AdS_3 \times S^3 \times {\cal M}^4$ where the 4-fold is $T^4$ or $K3$.  Let's act with those operators smeared with angle localized functions at some regularly spaced discrete set of points on the sphere, with maximal angular separation.  

It is a general theorem in QFT that a Green's function at finite total energy $E$ and fixed space-like separation, can be computed with arbitrary precision in a cut-off version of the QFT. Once the cutoff is at higher energy than $E$ and greater spatial precision than the separation between the points, the precision gets better as we increase the cutoff scale. 
So the only thing controversial about the last item of our assumptions is the finite dimension of the site Hilbert space.  It is of course an axiom of the Wilsonian approach to renormalization that many lattice theories can lead to the same continuum CFT.  It's well known for example that replacing non-compact continuum fields by compact lattice fields is an irrelevant deformation of the theory.  Compact lattice fields have infinite dimensional site Hilbert spaces, but the insistence on an {\it energy} cutoff in the Green's functions we're trying to compute implies that we can neglect all but a finite subspace of the ``angular momentum modes" at each site.  

On a more abstract level, a lattice cutoff with a certain spacing is morally equivalent to the statement that one can neglect higher derivatives of fields (higher dimensional descendants of primary operators) beyond a certain dimension.  In attempting to construct a lattice version of a CFT from its conformal bootstrap data, one might be tempted to throw away primaries beyond that same dimension.  The work on the Tensor Network Renormalization Group by Evenbly and Vidal\cite{tnrg} provides evidence that this is a correct strategy.  Their work constructed a sequence of embedding Hamiltonians of one layer of the network into the next larger one, which was obtained by insisting on local entanglement and applying the variational principle to exactly soluble critical lattice models.  Remarkably, the low lying spectrum of $K_0 + P_0$ for the CFT was reproduced by the spectrum of their embedding Hamiltonians for quite small lattices.  If one knew all the CFT data to begin with, one could build such a lattice TNRG with ease.  

If we accept that arbitrary Green's functions that lead to thermalization can be reproduced in a finite dimensional Hilbert space, the thermal density matrix itself lives in such a space.
We've thus argued that thermal states of the CFT, above the Hawking page transition for CFTs with an Einstein-Hilbert dual, can be modeled with arbitrary precision in a system with a finite total number of states.  This holds {\it a fortiori} for the meta-stable small black hole equilibria with Schwarzschild radii smaller than $R_{AdS}$.  Since the bounds\cite{bousso} on the entropy of general causal diamonds are motivated by the idea that higher entropy states would form black holes, there is no reason to associate an infinite dimensional operator algebra with any finite region of space time.   Our arguments depend on a technical conjecture about the approximation of finite energy states in CFTs by lattice theories with finite dimensional site Hilbert spaces.  This conjecture deserves more study.  If it is proven then the issue is closed.   

We should note that the general principles of renormalization theory, combined with the insight of\cite{susswit} that a finite volume cutoff on AdS space is equivalent to a UV cutoff on the boundary CFT, imply that there cannot be a unique definition of finite area causal diamonds that relies solely on the boundary data.  There are many cutoff models that have the same continuum limit.  One recent proposal is to use the TTbar deformation\cite{ttbar} of $1 + 1$ dimensional CFTs, and hypothetical higher dimensional generalizations of it, to define the cutoff.  This deformation produces complex eigenvalues of $K_0 + P_0$ above a certain level and the spectrum is truncated before this point, producing a finite dimensional Hilbert space.  The details of our argument would be very different for this case, and have not been fully worked out, but the conclusion is undoubtedly the same.  

The ambiguity of local physics reproducing fixed boundary data is familiar from ordinary quantum field theory, where the scattering matrix alone does not determine field correlation functions inside a Borchers class.  An analog of this in AdS/CFT has been found in the perturbative $1/N$ expansion of bulk fields by the authors of\cite{KL}. It is perhaps not unreasonable to conjecture that all cutoff schemes that preserve unitarity will obey the finiteness argument given above.  

Although our arguments do not apply directly to de Sitter space, given these remarks there really seems to be nothing to be gained from the post-modern suggestion that de Sitter space be described by a Type $II_1$ von Neumann algebra\cite{CLPW}.   The hyperfinite Type $II_1$ algebra can be approximated with arbitrary precision by a finite algebra of fermion creation and annihilation operators, so without even invoking any physics, we cannot distinguish this suggestion from one in which the Hilbert space is finite (although philosophically the Type $II_1$ algebra would have a mysterious commutant in a mysterious full algebra of all operators on a Hilbert space that described things not in the dS universe)\footnote{In a recent paper, Liu and Kolchmeyer\cite{LK} argued that once one takes into account quantum fluctuations of the position of the detector, the Type $II_1$ algebra is replaced by a direct integral of Type $I_{\infty}$ algebras. They argue that this is associated with the detector ``visiting all of the causally disjoint diamonds in dS space".  This disagrees with the conclusions of\cite{satbwf}, who showed that the wave function of the detector is spread over the horizon of a single diamond in a time of order $ R log (mR)$. This was interpreted as the return of the meta-stable isolated detector state to the empty dS equilibrium, which maximizes the entropy.}.  In fact we know that actual precision quantum mechanical measurements require a detector with many ``pointer q-bits", semi-classical variables whose values can be entangled with the q-bits of the system being measured.  The only way we know how to construct such systems in a dS universe is by making localized objects whose dynamics approximately obeys the rules of QFT.  The Schwarzschild de Sitter entropy formula shows us that a state containing such a localized object is a highly constrained state in the full dS universe.  Furthermore, the most generic state of that object occurs when it collapses into a black hole and becomes more or less useless as a detector.  So even no finite dimensional model of dS space is ever going to be testable with the precision that the mathematics pretends to give us.  

The conclusion from all of this is that many finite dimensional models of the Hilbert space of an asymptotically de Sitter space time are not distinguishable from each other by either mathematical arguments or any measurement that can, {\it in principle}, be made in such a universe.  Models with an infinite dimensional operator algebra add nothing to the study of de Sitter physics.  

\section{Meta-stable de Sitter vacua in String Theory?}

In this section we address a rather different question, motivated by the belief of many string theorists that string theory does not admit stable de Sitter universes. 
The first thing one should ask in attempting to address the question in the title of this section is what the stable state to which the meta-stable vacuum decays is supposed to be.  It's easy to see that it cannot be a supersymmetric Minkowski or AdS vacuum.  Although the task of finding convincing constructions of potentials in string theory that give rise to meta-stable stationary points with de Sitter geometry is difficult, it is easy to write down effective ${\cal N} = 1$, $d = 4$ super gravity models with such stationary points, which also have supersymmetric solutions with vanishing or negative c.c.  .

Starting with fields incoming from infinity we can arrange for scalar fields to collect in the vicinity of the meta-stable minimum over a region whose size is of order the putative dS radius.   The inevitable result is the formation of a black hole.  The black hole decays by Hawking radiation on a time scale much shorter than the Coleman Deluccia decay time of the dS space.  Moreover, the CDL decay of the dS state manifestly does not lead to the supersymmetric vacuum state.  In the negative c.c. case it leads to a Big Crunch.  Whatever the interpretation of these solutions of the gravitational field equations, they have nothing to do with a CFT that might be dual to the supersymmetric AdS stationary point. Entropy considerations suggest that if they represent anything, they are back and forth transitions between a dS state with a large finite entropy and a low entropy state represented by the crunch.  Neither is a state in the CFT Hilbert space.  For vanishing c.c.  the asymptotic space-time is not singular, but it is certainly neither supersymmetric nor Minkowski.  Speculations about it have appeared in\cite{susskindetal} but there is no definitive theoretical framework for these solutions.

Ignoring these questions of principle, we can ask what the practical utility of the meta-stable description of dS space might be, assuming that a solid framework for calculation can be found.  Here we are guided by our experience with meta-stable equilibrium states in all other parts of physics.  We do not need to have a detailed theory of the final states into which long lived nuclei or meta-stable atoms decay, in order to get a good computation of their spectra and of dynamical processes on time scales much shorter than the decay.  Thus, even if there is no sensible quantum theory of stable dS space, if a dS resonance exists, with the statistical properties (lifetimes) indicated by semi-classical calculations, then there must be an approximate finite dimensional model, valid on time scales short compared to the decay time, which gives an accurate account of physics on those time scales.  The most robust time coordinate to which we could possibly tie any model of physics that any civilization we can be in causal contact with could measure, is the proper time of the center of mass of some local group of galaxies, (preferably our own), over the interval from the creation of the earliest stars in that group until the whole system collapses into a black hole.  Any model of meta-stable dS space, which predicts decay times longer than that, will not be distinguishable by an experiment done by any entity that we can influence, from a finite dimensional model.  

  It thus appears that the suggestions of\cite{tb} and \cite{wf} are still the most promising paths towards a theory of quantum gravity applicable to our universe.  The original proposals made there, that the density matrix of the empty dS space was maximally uncertain, have proven to be wrong.  They've been replaced by the general ansatz of Carlip and Solodukhin\cite{CS}.  This has been tested in a variety of special cases\cite{VZ2}\cite{tbpdds}\cite{hezurek} and provides the basis for a quantitative alternative theory of the Cosmic Microwave Background\cite{satb} as well as possible tests in upcoming interferometer experiments\cite{interfer}.

\end{document}